\documentclass[11pt,twoside]{article}
\usepackage{asp2004}
\usepackage{epsfig}
\usepackage{graphics}
\usepackage{lscape}
\reversemarginpar

\begin{document}
\title{The Carnegie Supernova Project }

\author{Wendy L. Freedman (for the Carnegie Supernova Project)
\footnote{ Members of the Carnegie Supernova Project include: Ray
Carlberg, Alex Filippenko, Gaston Folatelli, Wendy Freedman, Mario
Hamuy, Weidong Li, Barry Madore, Nidia Morrell, Eric Persson, Mark
Phillips, Chris Pritchett, Nick Suntzeff, and Pamela Wyatt.}
} 
\affil{Carnegie Observatories, 813 Santa Barbara St., Pasadena, CA 91101, USA}

\begin{abstract}
The Carnegie Supernova Project (CSP) is aimed at providing an
independent measure of change in the the Hubble expansion as a
function of redshift, and setting constraints dark energy contribution
to the total energy content of the universe. Using type Ia supernovae
(SNIa), the CSP differs from other projects to date in its goal of
providing an I-band {\it restframe} Hubble diagram. The CSP is focused
on testing for and reducing systematic uncertainties, obtaining a
sample of multiwavelength observations of approximately 200 supernovae
over the redshift range 0 $<$ z $<$ 0.6. The $UBVRIYJHK_s$ data for
low-redshift supernova are intended to provide a database for the
determination of the Hubble constant, accurate K- and S-corrections,
comparison with theoretical models of supernovae, and for comparison
with the $RIYJ$ data of high-redshift supernovae. The goal is to
measure the evolution of the expansion rate, to characterize the
acceleration of the Universe, and constrain the equation of state, w,
to a precision and accuracy of 10
Type II SNae as independent distance indicators. Following an ongoing,
initial test period, the project will begin during the fall of
2004. Here an overview of the project is given, and some preliminary
results from the pilot program are presented.

\end{abstract}
\thispagestyle{plain}

\section{Introduction}

The evidence for an accelerating universe, with its implication for
the existence of a repulsive dark energy, is of profound significance
for particle physics and cosmology. Yet the explanation for the dark
energy remains a complete mystery.  There are at least two major
challenges to a theoretical understanding of the dark energy: 1) the
small magnitude of the dark energy component relative to its expected
value based on standard particle physics -- a discrepancy with the
observed value of 55 orders of magnitude or more, and 2) it appears
that we are living at an epoch when coincidentally the dynamics of the
expansion are only now becoming dominated by the dark energy.  Given
these immense challenges and the current lack of a physical
understanding of dark energy, further empirical characterization of
the evolution of the expansion rate of the Universe is clearly needed.

In general relativity, the expansion of the Universe, described in terms of the scale factor, a(t) can be written:


$$ \ddot{a} / a = -4\pi G \sum_i(\rho_i + 3 P_i) $$

\noindent where $\rho$ is the energy density and P is the pressure of
the various components (matter, radiation, dark energy) of the
Universe. Both energy and pressure govern the dynamics of the
universe. This equation allows for the possibility of both negative as
well as positive pressure, with a  negative pressure acting as an effective
repulsive gravity. Any component of the mass-energy density can be
parameterized by its ratio of pressure to energy density, w = P
/$\rho$.
          
In a universe with dark energy,  the expansion rate of the Universe is given by:
 
$$ H^2(z) / H^2_0 = [\Omega_m(1+z)^3+\Omega_\Lambda(1+z)^{3(1+w)}] $$

\noindent where  $\Omega_m$ and $\Omega_\Lambda$  represent the matter
and dark energy densities. For ordinary  matter w = 0, for radiation w
= 1/3, and for the cosmological constant (dark energy) w = -1.

There are two main observational approaches that currently provide
evidence for dark energy.  First are measurements of the Hubble
diagram using type Ia supernovae (SNIa), for which the best fit yields
a model with $\Omega_m \sim$ 0.3, and $\Omega_\Lambda \sim$ 0.7 (Riess
et al. 1998; 2004; Tonry et al. 2003; Perlmutter et al. 1999; Knop et
al. 2003). Second are measurements of the angular power spectrum of
the cosmic microwave background (CMB), which provide an independent
check on, and a consistent set of cosmological parameters as the SNIa
(Spergel et al. 2003, Page et al. 2003). The Wilkinson Microwave
Anisotropy Probe (WMAP) results, combined with measurements of
large-scale structure, yield results consistent with the type Ia
supernova measurements, with a matter density of about one third, and
the remaining two-thirds contribution from a dark energy component.

There are multiple advantages of using SNIa for measurements of
$\Omega_\Lambda$.  First, SNIa are luminous and can observed over a
wide redshift range. They offer a means of directly providing
measurements of a change in the expansion rate over time, and hence
for an acceleration of the Universe. (The CMB measurements provide
constraints on the energy density of an additional component such as
dark energy, but not on the acceleration of the Universe.) Second, the
dispersion in the SNIa Hubble diagram ($\sim$0.14 mag) is small enough
that the shape of the Hubble diagram can be used to separate
$\Omega_m$ and $\Omega_\Lambda$, independently of the nearby, local
calibration sample. Third, potential effects due to evolution,
chemical composition dependence, dust properties, and gravitational
lensing, can be empirically tested, calibrated, and corrected.

However, as ongoing and future supernova surveys yield larger sample
sizes, the statistical uncertainties will decrease further, and
systematics will dominate the total uncertainty. An increasing
challenge for ``precision measurements'' in cosmology, is
understanding and minimizing small systematic uncertainties, essential
for characterizing the nature of the dark energy.

\section{The Carnegie Supernova Project (CSP)}

The Carnegie Supernova Project (CSP) makes use of the unique resources
available to us at the Las Campanas Observatory (LCO): the  1-m Swope, 2.5-m
duPont,
and two 6.5-m Magellan telescopes, instrumented with CCDs and IR cameras and
CCD spectrographs. At low redshift, the goal is to provide
well-observed lightcurves from the near-ultraviolet to the near-IR
($UBVRIYJHK_s$)\footnote{The $Y$-band is centered at 1$\mu m$;
Hillenbrand et al. 2002.}.  An immediate result of this effort will be
a fundamental dataset on the photometric and spectroscopic sytematics
of both type Ia and II SN events.

The primary aim of the CSP is to establish a rest-frame I-band Hubble
diagram, while at the same time assembling an extensive database for low redshift
supernovae, useful for a variety of supernova studies.  For the Hubble
diagram, the I band represents the best compromise wavelength to work at; shorter restframe
passbands (UBV) have the advantage that they can be followed out to
higher redshifts, but they suffer larger systematic uncertainties.
The I passband offers important advantages for minimizing potential
systematic effects such as reddening and metallicity; however, the
objects cannot be observed to as great distances.  Hence, optical and
near-infrared observations remain quite complementary (see Figure 1).
However, by a redshift of 0.5, the differences amongst various
cosmological models are quite significant (that is, testing for a dark
energy component does not require observations to high redshifts).
The I-band restframe data will thus provide both a critical test of
the current shorter-wavelength restframe data, as well as an independent
measure of the dark energy component.

\begin{figure}[!hbt]
\begin{center} 
\psfig{file=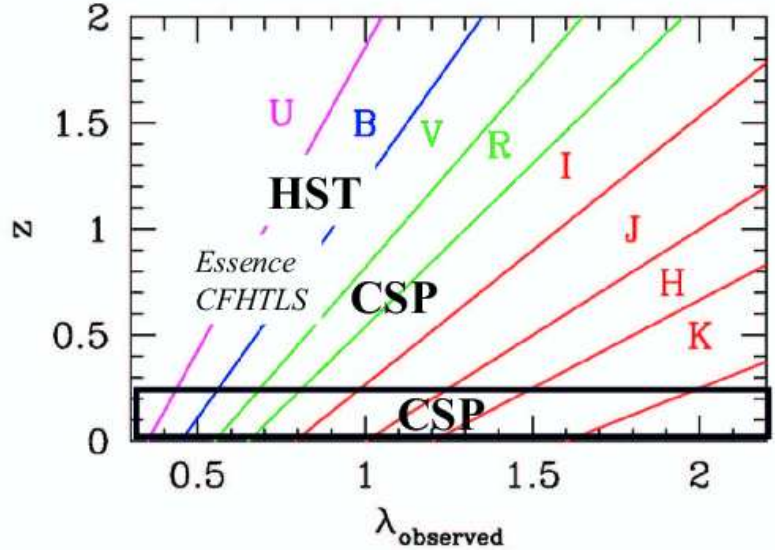,width=5in} 
\caption{Redshift and wavelength coverage. Diagonal lines indicate the
rest-frame wavelength passband; the redshift and wavelength at which
the object is observed can be read from the axes. At a redshift beyond
z=1, supernovae can still be observed at UB rest wavelengths using optical
CCDs. The redshift/wavelength domains of several ongoing supernova
projects are illustrated: HST at very high redshifts and bluer
bandpasses (z$>$1); ESSENCE/SNLS at more modest redshifts and optical
restframe coverage; the CSP with an extension to redder restframe
wavelengths using infrared arrays .  Locally the CSP will provide
extensive wavelength coverage from the ultraviolet through the
near-infrared.  }
\label{fig.zlambda}
\end{center} 
\end{figure}

Observations of low-redshift supernovae (z$<$0.2) are being carried
out using the 1-meter Swope and 2.5-meter Dupont telescopes. The
high-redshift (z$>$0.2) observations are being carried out at the
Baade 6.5-meter telescope, using the Persson Auxilliary Near-Infrared
Camera (PANIC).

We itemize here the many parts to and goals of the CSP:

1) To provide a reference data set with high signal-to-noise,
uniformly-calibrated photometry of type Ia supernovae, for comparison
with high-redshift supernovae for cosmology studies.

2) To provide bolometric light curves for comparison with theoretical
models for supernovae.

3) To obtain an independent measure of the Hubble constant based on infrared photometry.

4) To characterize in detail the nature of supernova light curves over
a range of wavelengths.

5) To investigate whether the light curves or spectra of type Ia supernovae 
differ with age of the progenitor.

6) To search for possible metallicity effects.

7) To study the connection of gamma-ray bursts and supernovae (in collaboration with Berger et al.)

8) To study nearby velocity flows -- deviations from the smooth Hubble expansion.

9) To improve the determination of photometric K-corrections

10) To improve S-corrections (spectral corrections).

11) To obtain independent estimates of the Hubble constant using type II supernovae.

12) To search for independent evidence of acceleration using type II supernovae.

\subsection{Low Redshift}

For the nearby sample, the goal is to obtain $UBVRIYJHK_s$ photometry
and optical spectroscopy for 125 low-redshift type Ia supernovae and
100 type II supernovae.  Photometric observations with a precision of
$\sim$0.03 mag will be obtained every 2-4 nights so that large gaps in
the light curves, common in supernova studies to date, can be
minimized.  The observations are being carried out from the time of
discovery through 50 days past maximum for SNIa, and through the
extended plateau phase for SNII.  An additional goal is to obtain optical
spectroscopy every 5-7 days from discovery through 40 days past
maximum.

The photometry obtained for many of the CSP supernovae will provide a
unique resource for improving the precision of these objects as
distance indicators, and for computing bolometric light curves for
comparison with theoretical models.  This nearby sample will serve as
a reference for the rest-frame $I$ and $Y$ light curves of the sample
of 120 high-redshift SNIa that we have begun to obtain. Furthermore,
the infrared photometry will be extremely valuable for independent
determinations of the Hubble constant as described below. It will also
be very useful for studying the nearby peculiar flows of galaxies out
to $\sim$10,000 km/s.

\medskip
\noindent
{\bf Near-Infrared Distances to Nearby Supernovae:}
Near-infrared observations offer the promise of improving the
precision of SNIa as cosmological standard candles for the
determination of H$_0$. Meikle (2000) and Elias et al.  (1985) have
noted the advantages of infrared photometry of supernovae, but the recent
availability of large-format infrared arrays now allows the full
potential of infrared observations to be exploited. Nearby supernova
distances can be calibrated, using the local Cepheid distance scale
to yield a value of H$_0$ at cosmologically
interesting distances.

Well-known advantages of infrared photometry are reduced
sensitivity to both reddening and metallicity. In addition, the $JHK$ data
for type Ia supernovae show peak absolute magnitudes in the near
infrared that are nearly constant, independent of decline rate (Meikle
2000; Krisciunas et al. 2004). The correlation between absolute
magnitude and decline rate is steepest in the $B$ band, and becomes
essentially flat at $H$ (see Figure 2). 

\begin{figure}[!hbt]
\begin{center} 
\psfig{file=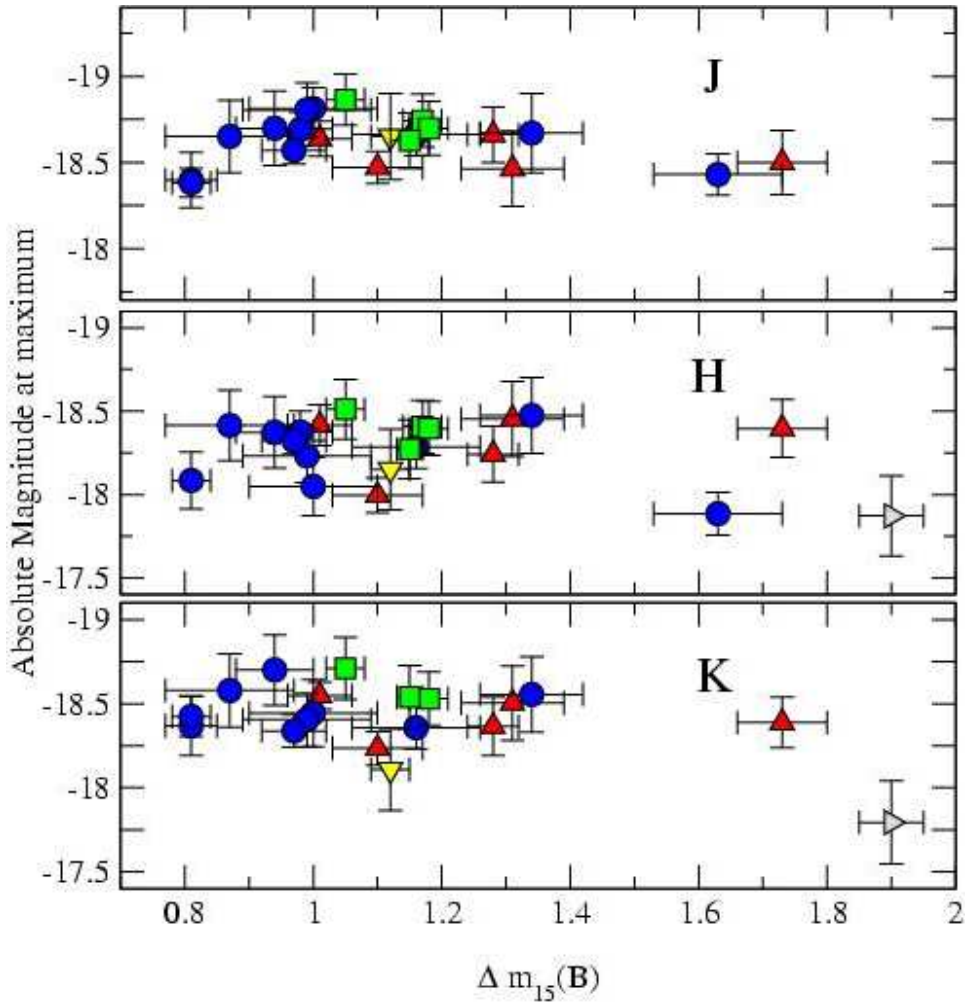,width=5in} 
\caption{Infrared     peak    magnitude     versus     decline    rate
relations.  Absolute  magnitudes  in  $JHK$ versus  the  decline  rate
parameter  $\Delta$m$_{15}(B)$  (Krisciunas et  al.  2004). Data  were
obtained at Las Campanas and Cerro Tololo Inter-American Observatory.}
\label{fig.deltamag}
\end{center} 
\end{figure}

We are also obtaining photometric and spectroscopic data on Type II
supernovae for an independent check on both the local distance scale,
and measurements of dark energy. Infrared measurements for SNII are
critical for reducing the systematic effects of reddening and
metallicity. Since SNII have young, massive progenitors, they are
found typically in regions with average higher extinctions than the lines
of sight to SNIa.

\subsection{ High Redshift}

To date, very few rest-frame $I$-band measurements have been obtained
for supernovae at high redshift. At $z$$\sim$0.25 this wavelength is
redshifted out of the reach of CCDs (thus requiring large-format
infrared arrays), while, in addition, the objects are faint (requiring
large telescopes).  Using the 6.5-meter Baade telescope, our goal over
the course of the next five years is
to observe a sample of $\sim$120 SNIa between $z$=0.2-0.6, obtaining
$YJ$ for each of the candidates, and $RI$ photometry for a subset of
the sample. These data will provide {\it rest-frame} $BVI$ photometry
for each supernova, yielding two colors, allowing accurate reddening
corrections to be determined.  We are obtaining 5-6 observations in $Y$ and $J$
covering the SN maximum and thereby allowing a firm measure of the peak
magnitude.  Because optical photometry is being obtained by the Legacy
and ESSENCE programs (see Section 3) for these supernovae, our
observations will afford a unique opportunity to ascertain the level
at which different photometric calibrations, K-corrections, and
reddening corrections impact the results.

HST NICMOS observations are available for a few high-redshift
supernovae (Riess et al. 2004), but due to practical limitations, the
bulk of measurements at high z are UV/optical restframe. Ultimately
observations from a future space mission (e.g., the Joint Dark Energy
Mission) may routinely obtain long-wavelength data for the
high-redshift supernovae (0.7 $<$ z $<$ 1.7). However, the
telescopes/instruments at Las Campanas offer a means to make
significant progress today from 0 $<$ z $<$ 0.7, and the CSP will provide a
means of eliminating reddening as a potential remaining source of
systematic error for this redshift range, while providing a fiducial
I-band comparison for future studies at higher redshifts.  If we live
in a universe where w=-1, then this redshift range is the one where
the cosmological effects of dark energy are manifest. At higher
redshifts, matter will dominate the expansion.

In late 2003 and early 2004, we obtained YJ photometry coverage for
eight supernovae with I-band magnitudes ranging from 21.3 to 24.0 mag,
and redshifts fairly uniformly distributed from 0.2 to 0.84. 



\subsection{Addressing Systematic Effects}

In an era of precision cosmology, where 10\% accuracy on the
measurement of w and 15\% on w' are desired goals, minimizing
the effects of systematic errors becomes the central issue to be
addressed.  Observations and careful study to date have demonstrated
that such systematic effects cannot explain away the observed
differences in supernova luminosities for the high- and low-redshift
samples. However, the requirement for increasing measurement accuracy --
and the lack of a detailed theoretical understanding of type Ia
supernovae, the current observations at restframe optical colors, the
difficulty of obtaining accurate K- and filter-corrections -- mean
that even previously small effects become important to characterize and eliminate.

\medskip
\noindent
{\bf Reddening: }
An advantage of longer-wavelength photometry 
is the decreased sensitivity to reddening. The ratio of total-to-selective 
absorption increases toward shorter wavelengths:

$$ R_\lambda =  A_\lambda / E(B - V) $$

\noindent
where the ratio of total-to-selective absorption, R, decreases from
$\sim$5 for the U-band, to $\sim$ 4 for the B-band, to 1.7 for the
I-band (Cardelli et al. 1989). Thus, the U-band absorption is a factor
of 3 greater at U than at I. In practice, this means that even for
very small reddenings, where E(B-V) $<$ 0.03, the corrections to the
restframe U-band magnitudes may be 0.15 mag; that is, comparable to the
cosmological effect being measured. Hence, at bluer restframe
wavelengths, the reddening corrections are more uncertain. One of the 
key goals of the CSP is to minimize the effects of reddening
in the Hubble diagram, and ensure that the {\it rest-frame} ($BVI$) bandpasses being
observed at low redshift match those for a sample at higher redshift, so that
reddening corrections can be applied in a uniform way.

\medskip
\noindent
{\bf Metallicity / Age:} Nearby SNIa occur in widely different stellar
environments, with varying ages and metallicities of their stellar
populations.  The multiwavelength nearby CSP dataset will provide an
excellent resource for addressing the question of whether there are
systematic differences due to metallicity or age of the progenitors
between the high- and low-redshift samples.

The effects of age and metallicity on the observed properties of SNIa
have been modeled by a number of investigators (e.g., H\"{o}flich et
al.  1998, Lentz et al. 2000).  These models suggest that
pre-explosion metallicity can have a significant effect on the
observed SNIa spectra.  For example, the models of Lentz et al. (2000)
indicate that scattering in the atmospheres is greatest in the U-band,
and decreases through the optical to infrared. However, predictions
from models to date have not yet converged on the sign or the
magnitude of such effects, and therefore, empirical constraints are
critical to minimize potential systematic effects in measurements of
the distances to SNIa.

To date, empirical searches for environmental dependences that might
correlate with the age of the supernova progenitor (host galaxy
morphology, color, position in the galaxy on supernova distances have
led to null results (e.g., Williams et al. 2003, Sullivan et
al. 2003).

\medskip
\noindent
{\bf Other Systematics:} Comparison of high- and low-redshift
supernovae for the measurement of cosmological parameters requires
accurate transformations of photometric bandpasses. The K-corrections
in use today are based on observations of a few low-redshift SNIa
whose overall spectral shapes from the ultraviolet through the
near-infrared are adjusted to match observed broad-band colors (e.g.,
Nugent et al. 2002).  Unfortunately, errors as large as 0.3 mag are
possible for some SNIa in the $I$ band (Strolger et al. 2002), and
large uncertainties remain in the U-band due to intrinsic variations
in the supernovae themselves, as well as due to the larger
extinctions.  One of the goals of the CSP is to increase the sample of
SNIa with a range of luminosities and decline rates, with
well-observed spectra and multiwavelength photometry for the purpose
of improving the K-corrections.

In addition, Stritzinger et al. (2002), have alarmingly found that the
peak magnitudes of supernovae can differ by up to 0.05 mag for data
taken at different telescopes, despite reducing the photometry to the
same local standards around the supernovae using the color terms
derived for each site and instrument.  Since a shift of 0.05 mag in
the $B-V$ color can introduce a 0.20 mag error in the extinction
corrected peak $B$ magnitude of a supernova, this is an additional
uncertainty that needs to be minimized when attempting to measure
cosmological parameters with higher precision than previous
measurements. A critical aspect of our low-redshift program is to
monitor our photometric systems in order to compute the spectral
corrections (S-corrections) required to bring the instrumental
magnitudes onto the standard photometric system, and decrease such
systematic effects.

\medskip
\noindent
%

\section{ Ongoing Supernova Searches }

{\bf Low Redshift:}
\medskip

\noindent
$\bullet$ LOTOSS: The source of low-redshift supernovae for the CSP is
primarily the Lick Observatory and Tenagra Observatory Supernova
Searches (LOTOSS).  LOTOSS is discovering supernovae over the redshift
range $z$=0.003-0.15, and obtaining $UBVRI$ light curves. Since 1998,
this survey has led to the discovery of about 400 supernovae.  The
survey efficiency has continued to increase with time so that over
half of the supernova discoveries have occured in the past 18 months.
In 2003, a collaboration between CSP and LOTOSS astronomers (Filippenko and 
Li) was begun, and the LOTOSS search fields
were shifted to include more galaxies in the southern hemisphere,
suitable for follow-up by the CSP. The collaboration allows the LOTOSS
group to concentrate its effort on the search, without the extra tax
in telescope time for follow-up, and the CSP is set up to carry out the
optical and near-infrared follow-up observations on the 1-m and 2.5-m
LCO telescopes.

\medskip
\noindent
{\bf High Redshift:}
\medskip

\noindent
Our high-redshift supernova targets  are coming from
the Supernova Legacy Survey (SNLS) and the   ESSENCE Project.
\medskip

\noindent
$\bullet$ The Legacy Survey is a Canadian/French collaboration, which
has been using the CFHT as of Feb. 2003 to obtain deep optical (ugriz)
images for 4 fields totaling 16 square degrees around the equator. The SNLS
is revisiting each field every second night during a 5-month campaign
per semester for the next 5 years. Their goal is to discover
2000 type I supernovae out to redshifts in excess of one, with 900 of
those having z $<$ 0.9.  They are  finding on average about 15
supernovae per month. In the week preceding CSP Magellan observations,
CSP and SNLS team members consult on candidates observed to be on the
rise, with spectra confirming types and providing redshifts. 
Carlberg and Pritchet of the Legacy project are CSP
collaborators.

\noindent
$\bullet$ ESSENCE is using the CTIO 4-m telescope to survey at VRI
wavelengths over the redshift range between $z$=0.15-0.75. They
revisit each field every second night during one 3-month (Oct-Dec)
campaign per year, and aim to produce optical light curves for 200
SNIa over 5 years (2002-2007).  The Essence Project
discoveries are also being made available to us in real time: targets,
classifications and redshifts. Suntzeff is the PI of this
project, as well as a collaborator on the CSP.

\section{Summary}

The Carnegie Supernova Project will go into full swing in the fall of
2004, with the goal of obtaining $UBVRIYJHK_s$ light curves and
optical spectroscopy for 225 low-redshift types I and II supernovae;
and $RIYJ$ photometry for 120 high-redshift type Ia supernovae over
the next 5 years.  This dataset will allow us to determine extinction
corrections, to constrain evolutionary effects due to age and
metallicity, and to minimize errors in supernova distances by
providing improved S-corrections.  These data will provide an
independent measure of the Hubble constant, and a restframe I-band
Hubble diagram. The ultimate goal of this research is to characterize
accurately the expansion history of the Universe, and to elucidate the
nature of the dark energy and its equation of state, w, to a precision
of better than 10\%.




%


\begin{references}


Cardelli, J. A., Clayton, G. C. \& Mathis, J. S., 1989, ApJ, 345, 245


Elias, J. H. et al. 1985, ApJ, 296, 379




Hillenbrand, L.A., Foster, J.B., Persson, S.E., \& Matthews, K., 2002, PASP, 114, 708

H\"{o}flich, P., Wheeler, J. C., \& Thielemann, F. K. 1998, ApJ, 495, 617

Knop, R. A., et al. 2003, ApJ, accepted, astroph-0309368


Krisciunas, K., Phillips, M. M. \& Suntzeff, N. B. 2004, astro-ph/0312626

Lentz, E. J., Baron, E., Branch, D., Hauschildt, P. H., \& Nugent, P. E. 2000, ApJ, 530, 966

Meikle, W. P. S., 2000, MNRAS, 314, 782

Nugent, P., Kim, A. \& Perlmutter, S.,  2002, PASP, 803

Page, L., et al. 2003, ApJ Suppl, 148, 175


Perlmutter, S., et al. 1999, ApJ, 517, 565


Riess, A. G., et al. 1998, AJ, 116, 1009


Riess, A. G., et al. 2004, ApJ, accepted, astroph-0402512


Spergel, D. et al. 2003, ApJ Suppl, 148,  175

Stritzinger, M. et al. 2002, AJ, 124, 2100

Strolger, L.-G. et al. 2002, AJ, 124, 2905

Sullivan, M., et al., 2003, MNRAS, 340, 1057


Tonry, J. L. et al. 2003, ApJ, 594, 1


Williams, B. F. et al., 2003, AJ, 126, 2608

\end{references}
\end{document}